\documentstyle[amsfonts,buckow]{article}

\begin{document}
\begin{flushright}
\small
IFUM-672-FT\\
hep-th/0012214
\normalsize
\end{flushright}
\def\titleline{
Dualities in $D=5$, $N=2$ Supergravity, Black Hole Entropy,
\newtitleline
and AdS Central Charges\footnote{Talk given at the RTN-workshop
{\it The Quantum Structure of Spacetime and the Geometric Nature of
Fundamental Interactions}, Berlin, October 2000}
}
\def\authors{
Dietmar Klemm\footnote{E-mail: dietmar.klemm@mi.infn.it}
}
\def\addresses{
Universit\`{a} degli Studi di Milano,
Dipartimento di Fisica,\\
Via Celoria 16,
20133 Milano,
Italy
}
\def\abstracttext{
The issue of microstate counting for general black holes
in $D=5$, $N=2$ supergravity coupled to vector multiplets
is discussed from various viewpoints. The statistical entropy
is computed for the near-extremal case by using the central
charge appearing in the asymptotic symmetry algebra of
$AdS_2$. Furthermore, we show that the considered supergravity
theory enjoys a duality invariance which connects electrically
charged black holes and magnetically charged black strings.
The near-horizon geometry of the latter turns out to be
$AdS_3 \times S^2$, which allows a microscopic calculation
of their entropy using the Brown-Henneaux central charges
in Cardy's formula. In both approaches we find perfect
agreement between statistical and thermodynamical entropy.
}
\large
\makefront
\section{Introduction}

The study of black hole solutions in $N=2$ five-dimensional supergravity
coupled to vector and hypermultiplets plays an important role in the
understanding of the non-perturbative structure of string and M-theory
\cite{chou,gaida}. In this setting the interplay between classical and quantum
results is exemplified at its best.

In this paper we consider general charged black holes of the $D=5$, $N=2$
theories, not necessarily those obtained from compactification
of eleven-dimensional supergravity on a Calabi-Yau threefold. The analysis
is simplified by the rich geometric structure of the $N=2$ theories. Black
hole solutions are given in terms of a rescaled cubic homogeneous
prepotential which defines very special geometry \cite{antoine}. In the
extremal BPS case, half of the vacuum supersymmetries are preserved, while
at the horizon supersymmetry is fully restored \cite{chamsabra2}.

Here we focus on the asymptotic symmetries of the near-horizon geometry of
the general near-extremal solution: the aim is the computation of the
entropy from a counting of microstates to be compared to the macroscopic,
thermodynamical entropy.

We will see that the calculation of the microscopic entropy of small
excitations above extremality is equivalent to a microstate counting for
certain black holes in two-dimensional anti-de~Sitter space. This can then
be done by using the central charge of the $AdS_2$
asymptotic symmetry algebra in Cardy's formula.

The main result however presented here
is an explicit duality transformation, which realizes
an invariance of the $N=2$ supergravity action \cite{cksz}. This duality turns the
$AdS_2 \times S^3$ near-horizon geometry of the extremal black hole
solution into $AdS_3 \times S^2$. The key point underlying the duality is
the fact that the three-sphere can be written as a Hopf fibration over the
base $S^2$. For $AdS_3$ the counting of microstates is performed using
the Brown-Henneaux central charges \cite{brown} in
Cardy's formula, and it is shown that this reproduces correctly the
Bekenstein-Hawking entropy.

In the case where the $D=5$, $N=2$ supergravity action is obtained by
Calabi-Yau (CY) compactification of M-theory, the considered duality
transformation, which maps electrically charged black holes onto
magnetically charged black strings, corresponds to the duality between M2
branes wrapping CY two-cycles and M5 branes wrapping CY four-cycles.
According to \cite{maldacena}, M-theory compactified on $AdS_{3}\times
S^{2}\times M$, where $M$ denotes some Calabi-Yau threefold, is dual to a
$(0,4)$ superconformal field theory living on an M5 brane wrapping some
holomorphic CY four-cycle. This fact has been used in \cite{vafa} to compute
the entropy of five-dimensional BPS black holes\footnote{The work in
\cite{vafa} includes as a special case also the results obtained
in \cite{stromvafa}.}. We stress that our method for microstate counting
applies to any near-extremal black hole in $N=2$, $D=5$ supergravity,
independent of whether it is obtained by CY compactification or not.

In section \ref{bhsection}, the black hole solutions of
$N=2$, $D=5$ supergravity coupled to vector multiplets are briefly reviewed.
We thereby focus on the
$STU$ model as a simple example, which nonetheless
retains all the interesting features of the general solutions.
In section \ref{ads2section} we compute the statistical
entropy of small excitations near extremality, using the $AdS_2$ central
charge \cite{cadmig}, and find perfect agreement with the Bekenstein-Hawking
entropy.
In section \ref{dualsection} we
construct the duality transformation for the supergravity action, and in
\ref{countingsection} we
finally perform the state counting, using the fact that
the near-horizon geometry of the dual solution includes an $AdS_3$ factor.
In this way, we obtain a microscopic entropy which agrees precisely with the
corresponding thermodynamical result.

\section{Black Holes in $N=2$, $D=5$ Supergravity}
\label{bhsection}

$N=2$, $D=5$ supergravity coupled to an arbitrary number $n$
of Maxwell supermultiplets was first considered in \cite{GST}.
In this theory, the scalar manifold can be regarded as a hypersurface
in an $(n+1)$-dimensional Riemannian space $\mathcal{R}$
with coordinates $X^I$. The equation of the hypersurface is
${\cal V}=1$ where ${\cal V}$, the prepotential, is a homogeneous
cubic polynomial in the coordinates of ${\cal R}$,
${\cal V}(X) = \frac{1}{6} C_{IJK}X^IX^JX^K$.
One can then parametrize the hypersurface in terms of the $n$ scalar
fields $\phi^i$
appearing in the vector multiplets, $X^I=X^I(\phi^i)$.

The bosonic part of the Lagrangian is given by
\begin{equation}
e^{-1}{\cal L}=\frac{1}{2}R-{\frac{1}{4}}G_{IJ}F_{\mu \nu }{}^{I}F^{\mu
\nu J}-\frac{1}{2}{\cal G}_{ij}\partial_{\mu}\phi^i\partial^{\mu}\phi^j
+{\frac{e^{-1}}{48}}\epsilon ^{\mu \nu \rho \sigma \lambda
}C_{IJK}F_{\mu \nu }^{I}F_{\rho \sigma}^{J}A_{\lambda}^{K}.
\label{generalact}
\end{equation}

The vector and scalar metric are completely encoded in the
function ${\cal V}(X)$,
\begin{equation}
G_{IJ} = -\frac12\partial_I\partial_J
\ln {\cal V}(X)|_{{\cal V}=1}, \qquad
{\cal G}_{ij} = G_{IJ}\partial_i X^I\partial_j X^J|_{{\cal V}=1},
\label{metric}
\end{equation}
where $\partial_{i}$ and $\partial_{I}$ refer, respectively, to partial
derivatives with respect to the scalar fields $\phi^{i}$ and
$X^I$. Note that for Calabi-Yau compactifications of M-theory,
$C_{IJK}$ denote the topological intersection numbers, ${\cal V}(X)$ represents the
intersection form, and $X^{I}$ and $X_{I}={\frac{1}{6}}C_{IJK}X^{J}X^{K}$ correspond,
respectively, to the size of the two- and
four-cycles of the Calabi-Yau threefold. In what follows, we will concentrate on
the $STU$ model \cite{chou,sabra1}, i.~e.~$X^0\equiv S$, $X^1\equiv T$, $X^2\equiv U$,
${\cal V}(X)=STU$. This model can be obtained by
compactification of heterotic string theory on $K_3 \times S^1$ \cite{antoniadis}.

The field equations following from the action (\ref{generalact}) admit the
non-extremal static black hole solutions \cite{bcs}
\begin{eqnarray}
ds^2 &=& -e^{-4V}fdt^{2}+e^{2V}(f^{-1}dr^{2}+r^{2}d\Omega _{3}^{2}),  
\nonumber \\
F_{rt}^{I} &=&-H_{I}^{-2}\partial _{r}\tilde{H}_{I}, \qquad 
X^{I} = H_{I}^{-1}e^{2V}, \label{staticbh}
\end{eqnarray}
where $d\Omega_3^2$ denotes the standard metric on the unit $S^3$.
The $H_I$ and $\tilde{H}_I$ are harmonic
functions,
\begin{equation}
H_I = 1+\frac{Q_{I}}{r^{2}}, \qquad \tilde{H}_{I}=1+\frac{\tilde{Q}_{I}}{r^{2}},
\end{equation}
where the $\tilde{Q}_{I}$ denote the physical electric charges.
$V$ and $f$ read
\begin{equation}
e^{2V}=(H_{0}H_{1}H_{2})^{1/3}, \qquad f=1-\frac{\mu }{r^{2}},
\end{equation}
with the nonextremality parameter $\mu$.
The physical charges are related to the $Q_{I}$
by the equations
\begin{equation}
Q_{I} = \frac{\mu}{2}\sinh\beta_I\tanh\frac{\beta_I}{2}, \qquad
\tilde{Q}_I = \frac{\mu}2\sinh\beta_I.
\end{equation}
The extremal (BPS) limit is reached when $\beta_I\rightarrow \infty$,
$\mu \rightarrow 0$, with $\mu \sinh \beta _{I}$ kept fixed.

For the ADM mass $M_{ADM}$, the Hawking temperature $T_H$, and the
Bekenstein-Hawking entropy $S_{BH}$, one obtains
\begin{equation}
M_{ADM} = \frac{\pi }{4G_{5}}(\sum_{I}Q_{I}+\frac{3}{2}\mu ), \qquad
T_H = \frac{\mu }{\pi \prod_{I}(\mu +Q_{I})^{1/2}}, \label{MADM}
\end{equation}
\begin{equation}
S_{BH} = \frac{A_{hor}}{4G_5} = \frac{\pi^2}{2G_5}\prod_I(\mu + Q_I)^{1/2}. \label{SBH}
\end{equation}
In the extremal case, the near-horizon geometry becomes $AdS_2 \times S^3$.

\section{Statistical Entropy from $AdS_2$ Central Charge}
\label{ads2section}

We would now like to use the near-horizon geometry $AdS_2 \times S^3$
to count the microstates which give rise to the black hole entropy (\ref{SBH}).
As we are mainly interested in the $AdS_2$ factor,
we perform a Kaluza-Klein reduction of the
$D=5$, $N=2$ supergravity action (\ref{generalact}) to two dimensions. As we
only consider nonrotating black holes carrying electric charge,
we can consistently truncate the Chern-Simons term in
(\ref{generalact}). The reduction ansatz for the metric is
\begin{equation}
ds^2 = \Phi^{-\frac{2}{3}}ds_2^2 + l_P^2\Phi^{\frac{2}{3}}d\Omega_3^2,
\end{equation}
where $\Phi$ denotes the dilaton
and $l_P$ is the Planck length in five dimensions.
In two dimensions, the field strenghts $F^I$
are proportional to the volume form and hence they can be integrated out.
In this way, one arrives at the two-dimensional action
\begin{equation}
I=\frac{\pi}{8}\int d^2x\sqrt{-g}\left[\Phi R + \frac{6}{l_P^2\Phi^{1/3}} - \Phi
{\cal G}_{ij}\partial_{\alpha}\phi^i\partial^{\alpha}\phi^j - \frac{G^{IJ}
\tilde{Q}_I\tilde{Q}_J}{l_P^6\Phi^{5/3}}\right].
\end{equation}
Let us now expand the nonextremal black hole solution (\ref{staticbh}) near extremality.
To this end, we introduce an expansion parameter $\epsilon$
($\epsilon \rightarrow 0$), and set
\begin{eqnarray}
t &=&\frac{\tilde{t}}{\epsilon },\qquad
r= \sqrt{\frac{2l_P^2\epsilon x}{(\tilde{Q}_0\tilde{Q}_1\tilde{Q}_2)^{(1/6)}} + \frac{\mu}
{2}}, \qquad
\mu =\mu _{0}\epsilon ,  \nonumber \\
\Phi &=& \frac{(\tilde{Q}_{0}\tilde{Q}_{1}\tilde{Q}_{2})^{1/2}}{l_P^3}
+ \frac{4}{\pi}\eta \qquad (\eta = {\cal O}(\epsilon)) ,\qquad \phi^i =
\tilde{Q}_i^{-1}(\tilde{Q}_{0}\tilde{Q}_{1}\tilde{Q}_2)^{1/3}
+ \epsilon\tilde{\phi}^i. \label{expansion}
\end{eqnarray}
One thus arrives at
\begin{equation}
ds^2 = -(\lambda^2 x^2 - a^2)d\tilde{t}^{2}+(\lambda^2x^2-a^{2})^{-1}dx^{2} \label{metr2d}
\end{equation}
for the two-dimensional metric, with $\lambda$ and $a$ given by
\begin{equation}
\lambda = \frac{2l_P}{(\tilde{Q}_{0}\tilde{Q}_{1}\tilde{Q}_{2})^{1/3}}, \qquad
a^{2} = \frac{\mu_0^2}{4l_P^2(\tilde{Q}_{0}\tilde{Q}_{1}\tilde{Q}_{2})^{1/3}}.
\end{equation}
The action at lowest order in the expansion parameter $\epsilon$ reads
\begin{equation}
I=\frac{1}{2}\int d^{2}x\sqrt{-g}\eta [R + 2\lambda^2],
\label{JT}
\end{equation}
so the leading order is governed by the Jackiw-Teitelboim model.
(\ref{metr2d}), together with the linear dilaton
\begin{equation}
\eta = \eta _{0}\lambda x, \qquad
\eta_0 = \frac{\Omega \epsilon }{16\pi l_P^2}(\tilde{Q}_0\tilde{Q}_1\tilde{Q}_2)^{2/3}
\sum_I\tilde{Q}_I^{-1},
\end{equation}
represents a black hole solution of this model \cite{cadmig}, with mass,
thermodynamical entropy and temperature given by
\begin{equation}
M_{(2)} = \frac12\eta_0a^2\lambda, \qquad S_{(2)} = 2\pi\eta_{hor} = 2\pi\eta_0a, \qquad
T_{(2)} = \frac{a\lambda}{2\pi}. \label{MS2}
\end{equation}
This black hole spacetime has constant curvature, i.~e.~it is locally $AdS_2$.
Now it is known that the asymptotic symmetries of two-dimensional
anti-de~Sitter space form a Virasoro algebra \cite{cadmig}, similar to the
case of $AdS_3$, where one has two copies of Virasoro algebras as
asymptotic symmetries \cite{brown}. This algebra was shown to have a
central charge $c=12\eta_0$ \cite{central}.

Expanding the ADM mass $M_{ADM}$ (\ref{MADM}) and Bekenstein-Hawking entropy $S_{BH}$
(\ref{SBH}) of the black hole (\ref{staticbh}) in five dimensions for $\mu \rightarrow 0$,
one obtains that small excitations above extremality have the energy and entropy
\begin{equation}
\Delta M_{ADM} = \frac{\pi\mu^2}{32l_P^3}\sum_I\tilde{Q}_{I}^{-1}, \qquad
\Delta S_{BH} = \frac{\pi^2\mu}{8l_P^3}(\tilde{Q}_0\tilde{Q}_1\tilde{Q}_2)^{1/2}\sum_I
\tilde{Q}_I^{-1}.
\end{equation}
Comparing this with the two-dimensional results (\ref{MS2}), one finds $\Delta S_{BH}
=S_{(2)}$ and $\Delta M_{ADM}=\epsilon M_{(2)}$. The factor $\epsilon$ appearing in the
relation between the two masses stems from the
fact that $M_{ADM}$ was computed with respect to the Killing vector $\partial_t$, whereas
$M_{(2)}$ is related to $\partial_{\tilde{t}}=\epsilon\partial_t$.
This means that up to these normalizations the
five- and two-dimensional energies and entropies match.

Let us now compute the statistical entropy.
Inserting the conformal weight $L_0 = M_{(2)}/\lambda$ together with
the central charge in Cardy's
formula $S_{stat} = 2\pi\sqrt{cL_0/6}$
yields a statistical entropy
which agrees precisely with the thermodynamical entropy $\Delta S_{BH}$ of the small
excitations above extremality.

\section{Duality Invariance of the Supergravity Action}
\label{dualsection}

In this section we will show that in presence of a Killing vector field $\partial_z$,
the supergravity action (\ref{generalact}) is invariant under
a certain generalization of T-duality\footnote{In what follows, we consistently truncate
the Chern-Simons term. One can easily generalize the discussion below to
nonvanishing CS term. This results in a $\theta$ term in four dimensions,
which does not spoil the considered duality invariance.}. The key
observation is then that the three sphere $S^3$ appearing in the black
hole geometry can be written as a Hopf fibration, i.~e.~as an $S^1$ bundle
over ${\mathbb C} P^1\approx S^2$. Performing then a duality transformation
along the Hopf fibre untwists the $S^3$, and transforms the electrically
charged black hole into a magnetically charged black string, which has $AdS_3\times S^2$
as near-horizon limit in the extremal case.

To begin with, we reduce the action (\ref{generalact}) to four dimensions,
using the usual Kaluza-Klein reduction ansatz for the five-dimensional
metric,
\begin{equation}
ds^{2}=e^{k/\sqrt{3}}ds_{4}^{2}+e^{-2k/\sqrt{3}}(dz + {\cal A}_{\alpha}
dx^{\alpha })^{2}, \label{KK}
\end{equation}
where $k$ denotes the dilaton, and early greek indices $\alpha,\beta,\ldots$
refer to four-dimensional spacetime.
One thus arrives at the
four-dimensional action
\begin{equation}
I_{4}=\frac{L}{16\pi G_{5}}\int d^{4}x\sqrt{-g_{4}}\left[ R_{4}-\frac{1}{2}%
(\nabla k)^{2}-\frac{1}{4}e^{-\sqrt{3}k}{\cal F}^{2}-\frac{1}{2}e^{-k/%
\sqrt{3}}F^{2}-{\cal G}_{ij}\partial_{\alpha}\phi^i\partial^{\alpha}\phi^j\right],
\label{action4d}
\end{equation}
where $L$ denotes the length of the circle parametrized by $z$,
${\cal F}$
is the field strength associated to the Kaluza-Klein vector potential
${\cal A}$, and
\begin{equation}
{\cal F}^{2} = {\cal F}_{\alpha\beta}{\cal F}^{\alpha\beta},
\qquad F^{2}=G_{IJ}F_{\alpha \beta }^{I}F^{J\alpha \beta }.
\end{equation}
We now dualize both ${\cal F}$ and $F^I$, which yields
\begin{eqnarray}
I_4 &=&\frac{L}{16\pi G_{5}}\int d^{4}x\sqrt{-g_{4}}[R_{4} -
\frac12 (\nabla k)^{2}-\frac{1}{4}e^{\sqrt{3}k}(^{\star }{\cal F})^2
\nonumber \\
&&- \frac{1}{2}e^{k/\sqrt{3}}\frac{1}{4}G^{IJ}\,^{\star }F_{I\alpha
\beta }\,^{\star}F_{J}^{\alpha \beta }-{\cal G}_{ij}\partial_{\alpha}\phi^i
\partial^{\alpha}\phi^j], \label{dualaction4d}
\end{eqnarray}
where we defined
\begin{equation}
^{\star }{\cal F}_{\alpha\beta} = \frac{1}{2}e^{-\sqrt{3}k}\epsilon
_{\alpha \beta \gamma \delta }{\cal F}^{\gamma \delta }, \qquad
^{\star }F_{I\alpha\beta} = e^{-k/\sqrt{3}}G_{IJ}\epsilon _{\alpha \beta
\gamma \delta }F^{J\gamma\delta}.
\end{equation}
Comparing (\ref{dualaction4d}) with (\ref{action4d}), we observe that the
gravitational and gauge field parts of the four-dimensional action, as well
as the dilaton kinetic energy, are invariant under the ${\mathbb Z}_4$
transformation
\begin{equation}
k\rightarrow -k,\qquad {\cal F}_{\alpha \beta }\rightarrow \,^{\star}{\cal F}_{\alpha
\beta}, \qquad F_{\alpha \beta }^{I}\rightarrow
\,^{\star }F_{I\alpha \beta },\qquad G_{IJ}\rightarrow \frac{1}{4}G^{IJ}.
\label{duality}
\end{equation}
The ${\mathbb Z}_4$ is actually a subgroup of the usual symplectic $Sp(2m+2,{\mathbb R})$
duality group \cite{craps} of $D=4$, $N=2$ supergravity (coupled to
$m$ vector multiplets).
Note that the transformation $G_{IJ}\rightarrow G^{IJ}/4$ means that
\begin{equation}
X^{I} \rightarrow 3X_{I}=\frac{1}{2}C_{IJK}X^{J}X^{K}, \qquad
X_{I} \rightarrow \frac{1}{3}X^{I}, \label{dualcoord}
\end{equation}
so essentially the special coordinates go over into their duals.
As (\ref{dualcoord}) does not change the kinetic term of the
scalar fields, (\ref{duality}), (\ref{dualcoord}) represent in fact a
duality invariance of the four-dimensional action (\ref{action4d}). In the
special case of the $STU=1$ model, (\ref{dualcoord}) implies that the moduli
$\phi^i$ go over into their inverse,
$\phi^i\rightarrow 1/\phi^i$.
We now wish to apply the duality (\ref{duality}), (\ref{dualcoord}) to the
black hole solution (\ref{staticbh}). To this end, we consider the $S^{3}$
as an $S^{1}$ bundle over $S^{2}$, and write for its metric
\begin{equation}
d\Omega_3^2 = \frac{1}{4}\left[ d\vartheta ^{2}+\sin ^{2}\vartheta d\varphi
^{2}+(d\zeta +\cos \vartheta d\varphi )^{2}\right] ,
\end{equation}
where $\zeta $ ($0\leq \zeta \leq 4\pi $) parametrizes the $S^{1}$ fibre.
Introducing the coordinate $z=\lambda \zeta $, where $\lambda $ denotes an
arbitrary length scale, one can write the 5d metric in the KK form 
(\ref{KK}), where
\begin{eqnarray}
ds_{4}^{2} &=&\frac{re^{V}}{2\lambda }\left[
-e^{-4V}fdt^{2}+e^{2V}f^{-1}dr^{2}+e^{2V}\frac{r^{2}}{4}(d\vartheta
^{2}+\sin ^{2}\vartheta d\varphi ^{2})\right], \nonumber \\
e^{-k/\sqrt{3}} &=&\frac{re^{V}}{2\lambda }, \qquad
{\cal A} = \lambda\cos\vartheta d\varphi.
\end{eqnarray}
We now dualize in 4d according to (\ref{duality}), and then relift
the solution to five dimensions. This yields the configuration
\begin{eqnarray}
ds^{2} &=&e^{-2V}\left[ \frac{\mu }{4\lambda ^{2}}dt^{2}+2dzdt+\frac{%
4\lambda ^{2}}{r^{2}}dz^{2}\right] +\frac{r^{2}}{4\lambda ^{2}}e^{4V}\left[
f^{-1}dr^{2}+\frac{r^{2}}{4}d\Omega _{2}^{2}\right] , \nonumber \\
F_{\vartheta \varphi }^{I} &=&\frac{\tilde{Q}_{I}}{4\lambda }\sin\vartheta,
\qquad X^I = H_I e^{-2V}. \label{magnblstr}
\end{eqnarray}
One effect of the duality transformation is thus the untwisting of the Hopf
fibration\footnote{%
The fact that Hopf bundles can be untwisted by T-dualities was observed in
\cite{dlp}. The idea of untwisting and twisting fibres
to relate strings and black holes, and thus to gain new insights into
black hole microscopics, was also explored in \cite{cvetic2}.}.

One can further simplify (\ref{magnblstr}) by an $SL(2,{\mathbb R})$ transformation
\begin{equation}
\left(
\begin{array}{c}
t^{\prime } \\
z^{\prime }
\end{array}
\right) =\left(
\begin{array}{cc}
0 & -\frac{2\lambda }{\sqrt{\mu }} \\
\frac{\sqrt{\mu }}{2\lambda } & \frac{2\lambda }{\sqrt{\mu }}
\end{array}
\right) \left(
\begin{array}{c}
t \\
z
\end{array}
\right) .  \label{SL2R}
\end{equation}
Introducing also the new radial coordinate $\rho =r^{2}/(4\lambda )$, we
then get for the metric
\begin{equation}
ds^{2}=e^{-2V}(-fdt^{\prime }{}^{2}+dz^{\prime }{}^{2})+e^{4V}(f^{-1}d\rho
^{2}+\rho ^{2}d\Omega _{2}^{2}).  \label{nonextrblstr}
\end{equation}
(\ref{nonextrblstr}), together with the gauge and scalar fields given in (%
\ref{magnblstr}), represents a nonextremal generalization of the
supersymmetric magnetic black string found in \cite{chamsabra2}. The duality
(\ref{duality}) thus maps electrically charged black holes onto
magnetically charged black strings.

\section{Microstate Counting from $AdS_3$ Gravity}
\label{countingsection}

We now want to use the near-horizon geometry of the dual solution (\ref{nonextrblstr})
to count the microstates giving rise to the
Bekenstein-Hawking entropy. In \cite{chamsabra2} it was shown that in the
extremal case, the geometry becomes $AdS_3\times S^2$ near the event
horizon. The idea is now to use the central charge of $AdS_3$ gravity 
\cite{brown} in Cardy's formula, in order to compute the statistical entropy,
like it was done by Strominger \cite{strominger1} for the BTZ black
hole\footnote{Cf.~also \cite{cvetic3}, where similar computations for black
strings in six dimensions with $BTZ \times S^3$ near-horizon geometry
were performed.}.
As only the $AdS_{3}$ part is relevant, we would like to reduce the
supergravity action from five to three dimensions. To this end, we first
Hodge-dualize the magnetic two-form field strength in (\ref{magnblstr}).
For the solution under consideration, the field strengths $H_I$ dual to the
$F^I$ do not depend on
the coordinates of the internal $S^2$. Furthermore, in 3d the three-forms 
$H_I$ are proportional to the volume form and can be integrated out. For
the metric, we use the reduction ansatz
\begin{equation}
ds^2 = \Phi^{-1}ds_3^2 + l_P^2\Phi^2d\Omega_2^2,
\end{equation}
where $d\Omega_2^2$ denotes the standard
metric on the unit $S^2$. This gives the reduced action
\begin{equation}
I=\frac{1}{4l_P}\int d^{3}x\sqrt{-g}\Phi^{\frac{3}{2}}\left[R
+\frac{2}{l_P^{2}\Phi^{3}}-\frac{3}{2\Phi ^{2}}(\nabla \Phi )^{2}-\frac{%
G_{IJ}P^{I}P^{J}}{\Phi ^{5}l_P^{4}}-{\cal G}_{ij}\partial _{\alpha }\phi
^{i}\partial ^{\alpha }\phi ^{j}\right], \label{action3d}
\end{equation}
where we introduced the magnetic charges
$P^I = \tilde{Q}_I/(4\lambda)$
of the black string (\ref{magnblstr}).
The idea is now to expand the 3d metric $ds_3^2$ near the horizon and near extremality.
This can be done by setting
\begin{equation}
t^{\prime }=\frac{t^{\prime \prime }}{\sqrt{\epsilon}}(2\lambda 
)^{4}\sqrt{\frac{l_P}{\mu_0\lambda\tilde{Q}_0\tilde{Q}_1\tilde{Q}_2}}, \qquad
z^{\prime }=\frac{z^{\prime \prime }}{\sqrt{\epsilon }}\frac{(2\lambda 
)^{2}%
}{\sqrt{\mu _{0}}},\qquad \rho =\epsilon \tilde{r}^{2}\frac{\mu_0l_P}{(2\lambda)^4},
\qquad \mu =\mu_0\epsilon,
\end{equation}
and taking the limit $\epsilon \rightarrow 0$. This leads to the metric
\begin{equation}
ds_3^2 = -\frac{\tilde{r}^{2}-\tilde{r}_{+}^{2}}{l_{eff}^{2}}%
dt^{\prime \prime }{}^{2}+\tilde{r}^{2}dz^{\prime \prime }{}^{2}+\frac{%
l_{eff}^{2}d\tilde{r}^{2}}{\tilde{r}^{2}-\tilde{r}_{+}^{2}},  \label{BTZ}
\end{equation}
where we introduced
\begin{equation}
\tilde{r}_{+}^{2} = \frac{4\lambda^3}{l_P}, \qquad
l_{eff}^{2} = \frac{\tilde{Q}_{0}\tilde{Q}_{1}\tilde{Q}_{2}}{16l_P\lambda 
^{3}}.
\end{equation}
We recognize (\ref{BTZ}) as the BTZ black hole, with event
horizon at $\tilde{r}=\tilde{r}_{+}$.
$\Lambda_{eff}=-1/l_{eff}^{2}$ is the effective cosmological constant. The effective
3d Newton constant can be read off from the action (\ref{action3d}),
yielding
\begin{equation}
\frac{1}{16\pi G_{eff}}=\frac{1}{4l_P}\Phi_{hor}^{3/2},
\end{equation}
where the subscript indicates that the dilaton $\Phi$ is to be evaluated at
the horizon.
The BTZ black hole mass is given by
\begin{equation}
M_{(3)}=\frac{\lambda^3}{2l_P G_{eff}l_{eff}^2}.
\end{equation}
We can now apply Strominger's counting of microstates \cite{strominger1} to
reproduce the Bekenstein-Hawking entropy. To this end, one first observes
that the central charge appearing in the asymptotic symmetry algebra of $%
AdS_{3}$ \cite{brown} in our case reads
$c = 3l_{eff}/(2G_{eff})$.
Furthermore, we have the relations
\begin{equation}
M_{(3)} = \frac{1}{l_{eff}}(L_{0}+\bar{L}_{0}), \qquad
J = L_{0}-\bar{L}_{0} = 0
\end{equation}
for the mass and angular momentum. We then obtain from Cardy's formula
a statistical entropy which coincides precisely with the
thermodynamical entropy (\ref{SBH}) of
the 5d black hole (\ref{staticbh}).

\vskip0.5cm
\noindent
{\large \bf Acknowledgements}

\smallskip
\noindent
I would like to thank my collaborators on the work which was the
subject of this talk, S.~Cacciatori, W.~A.~Sabra and D.~Zanon.
This work has been partly supported
by MURST and by the European Commission RTN program
HPRN-CT-2000-00131, in
which the author is associated to the University of Torino.

\end{document}